\documentclass[british,english]{iopart}
\usepackage[T1]{fontenc}
\usepackage[latin9]{inputenc}
\usepackage{amstext}
\usepackage{amssymb}
\usepackage{graphicx}
\usepackage{esint}

\makeatletter

\DeclareRobustCommand{\greektext}{%
  \fontencoding{LGR}\selectfont\def\encodingdefault{LGR}}
\DeclareRobustCommand{\textgreek}[1]{\leavevmode{\greektext #1}}
\DeclareFontEncoding{LGR}{}{}
\DeclareTextSymbol{\~}{LGR}{126}
\newcommand{\lyxdot}{.}

\usepackage{iopams}
\usepackage{setstack}

\makeatother

\usepackage{babel}
\begin{document}

\title{Coherent topological defect dynamics and collective modes in superconductors
and electronic crystals.}

\author{D. Mihailovic$^{1,2}$ T.Mertelj$^{1}$, V.V.Kabanov$^{1}$ and S.Brazovskii$^{3}$. }

\address{$^{1}$Jozef Stefan Institute, Jamova 39, 1000 Ljubljana, Slovenia }

\address{$^{2}$$ $CENN Nanocenter, Jamova 39, 1000 Ljubljana, Slovenia}

\address{$^{3}$ LPTMS-CNRS, U. Paris-Sud, bat. 100, Cite Université, 91405
Orsay, France}
\begin{abstract}
The control of condensed matter systems out of equilibrium by laser
pulses allows us to investigate the system trajectories through symmetry-breaking
phase transitions. Thus the evolution of both collective modes and
single particle excitations can be followed through diverse phase
transitions with femtosecond resolution. Here we present experimental
observations of the order parameter trajectory in the normal$\rightarrow$superconductor
transition and charge-density wave ordering transitions. Of particular
interest is the coherent evolution of topological defects forming
during the transition via the Kibble-Zurek mechanism, which appears
to be measurable in optical pump probe experiments. Experiments on
CDW systems reveal some new phenomena, such as coherent oscillations
of the order parameter, the creation and emission of \foreignlanguage{british}{dispersive}
amplitudon modes upon the annihilation of topological defects, and
mixing with weakly coupled finite-frequency (massive) bosons. 
\end{abstract}
\maketitle

\section{Introduction. \label{sec:Introduction.}}

The idea that non equilibrium phenomena in natural systems can be
described in terms of a temporal evolution of a non-linear Landau
expansion of the free energy has lead to theories as wide afield as
elementary particles to cosmology. In condensed matter systems, where
the idea originates, the opportunities for the study of the \foreignlanguage{british}{behaviour}
of nonlinear systems abound. However, typically, the study of these
systems has been limited to near-equilibrium situations, where the
system evolves slowly through the transition. Yet both particle physics
experiments and cosmology distinctly deal with the temporal evolution
of highly non-equilibrium systems. Typically one detects the decay
products well after the decay itself, in the aftermath of the symmetry-breaking
transition (SBT). Collisions of elementary particles, the subsequent
creation of a high-symmetry intermediate state and its decay via symmetry-breaking
interactions takes place on timescales which are well beyond the resolution
of our current technology. Cosmological experiments being out of reach,
experimental cosmological studies of the aftermath of the Big Bang
reveal very complex \foreignlanguage{british}{behaviour} in its aftermath,
still beyond current theoretical understanding. As a result we have
little insight into the non-equilibrium conditions close to the critical
time $t_{c}$ of the SBT when the creation of the new state is taking
place. 

In condensed matter systems the intrinsic dynamics of elementary excitations
occur on timescales which are becoming accessible with current femtosecond
laser technology. Particularly the normal-to-superconducting state
transition and charge-density wave transition are of fundamental interest,
as examples with different symmetry properties. Thus, typical single
particle (quasiparticle) relaxation times in high-temperature superconductors
are on timescales of $10^{-12}$ s \cite{Kabanov-1}, while electronic
energy relaxation occurs on timescales of $10^{-12}-10^{-13}$s \cite{Gadermaier}.
Many body collective states have similar characteristic timescales.
The so-called Ginzburg Landau (GL) time $\tau_{GL}$, \cite{Schmid}
which enters into the time-dependent GL equation is $\tau_{GL}\simeq1/\Delta_{s}$
where $\Delta_{s}$ is the superconducting gap. For high-temperature
superconductors, typical \foreignlanguage{british}{estimates} give
$\tau_{GL}\sim10^{-13}-10^{-12}$ s. In superconducting systems, the
collective mode is overdamped, which means that it is difficult to
discern the relaxation of the collective mode from single particle
excitations. On the other hand, the characteristic timescale of collective
mode in charge density wave systems is typically around $2\pi/\omega_{AM}\simeq0.5\times10^{-12}$
s \cite{KMO,Yusupov2-1}. Laser pulses can currently be created with
sub femtosecond duration, so ample resolution is available for the
study of the coherent evolution of collective modes in superconducting
and charge-density-wave (CDW) systems. Moreover, by studying the concurrent
evolution of the single particle spectrum, it may be possible to investigate
the transient state of the underlying vacuum.

Following the suggestion by Zurek\cite{Zurek} of laboratory experiments
to test Kibble's cosmological model \cite{Kibble}, \textquotedblleft{}system
quench\textquotedblright{} experiments were performed in a number
of different systems, including superconductors \cite{Monaco,Golubchik},
Bose-Einstein (BE) condensates within a trapped atomic gas \cite{Weiler}
and polariton BE condensates \cite{Roumpos}. Experiments on the Kibble-Zurek
(KZ) mechanism so far have concentrated on the statistical analysis
of topological defects left behind by the SBT, on timescales long
compared to the intrinsic GL time \textgreek{t}$_{GL}$. Quench rates
in these systems were relatively slow, the fastest being around $10^{8}$K/s
\cite{Golubchik}. 

In standard optical Pump-probe \emph{(P-p)} techniques (including
THz probe) the response is related to the modulation of the dielectric
constant (reflectivity or absorbance), which can have a number of
contributions in electronic systems such as superconductors and CDWs
which are in one way or another related to the order parameter $\psi$:
(1) the response due to hot carrier energy relaxation, (2) the quasiparticle
(QP) recombination across the SC or CDW gap and (3) QP recombination
across a pseudo-gap which is often present in these systems and  (4)
coherent phonon oscillations due to displacive excitation or impulsive
Raman excitation. The coherent phonon oscillations cannot be distinguished
from the oscillations of the order parameter with \emph{P-p} spectroscopy.
Moreover, different phonon modes are coupled to $\psi$ to various
degrees, and the dephasing is different for each mode, thus adding
further to the complexity of the response from which it is difficult
to extract the order parameter. We show that using a 3-pulse technique,
these responses can be isolated to some extent, allowing us to study
the coherent evolution of the order parameter with time though an
SBT with very high temporal resolution. The method allows us to study
the interactions of the collective mode of the CDW with other modes
of different symmetry as well as annihilation of topological defects,
revealing new finite frequency dispersive Higgs-like field excitations
released as a result of domain wall annihilation.

\section{Quench experiments with a multiple pulse all-optical technique.\label{sec:Quench-experiments-with}}

The principle behind the technique involves the use of a multiple
pulse trains to control the system and measure the time-response of
the order parameter. A schematic diagram of the pulse sequence is
shown in Fig. 1.
\begin{figure}
\includegraphics[width=0.6\columnwidth]{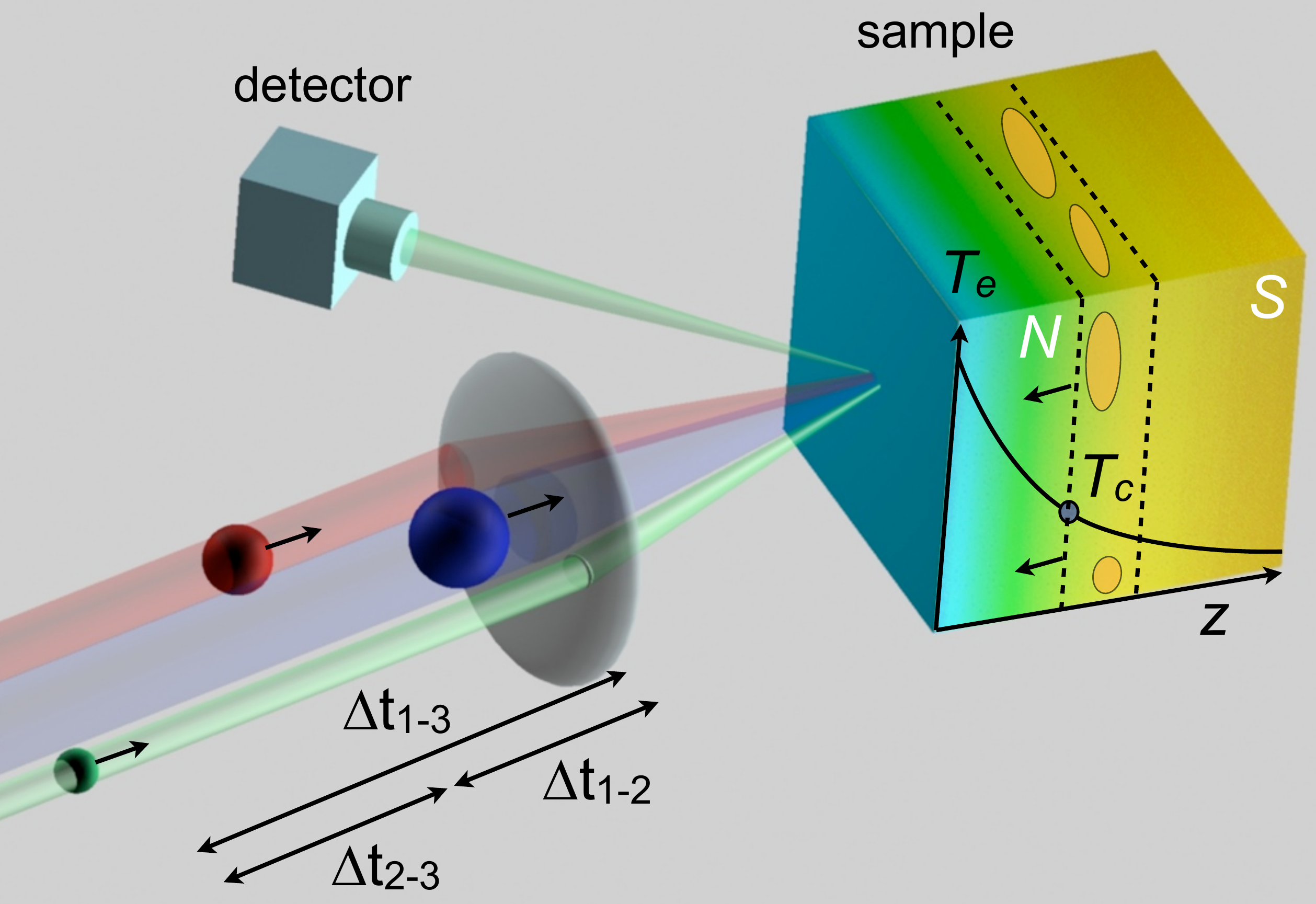}

\caption{A schematic diagram of the three-pulse sequence and evolution of the
superconducting order in a superconductor. The different time delays
are indicated. The ratio of intensities of the D (blue ball), P(red
ball) and p(green ball) laser pulses are typically 500:10:1 respectively.\label{fig:A-schematic-diagram}
The electronic temperature decays into the sample as a result of the
inhomogeneous excitation by laser light. The position of a normal/superconducting
(N/S) state front, given by $T_{e}=T_{c}$ is rapidly moving towards
the surface, leaving vortices in its wake. In a CDW system, the D
pulse forces the system into the high-symmetry state, whereupon it
undergoes a SBT as it cools. In the process, topological defects are
created as a result of the inhomogeneity of the laser field.}
 
\end{figure}
A strong laser destruction (\emph{D}) pulse - whose intensity is carefully
adjusted to cause a perturbation of appropriate magnitude - is used
to rapidly transfer the system from an ordered (broken symmetry) state
to the disordered (high-symmetry) state, whereafter the system reverts
back to equilibrium through the SBT. The state of the system at any
given time after the \emph{D} pulse is determined by standard pump-probe
spectroscopy sequence, which involves first exciting the system by
an additional weak perturbing \emph{P} pulse delayed by $\Delta t_{1-2}$
with respect to the \emph{D} pulse, and subsequently measuring the
resulting change of reflectivity $\Delta R/R$ of the sample by a
weak probe (\emph{p}) pulse (see Fig. \ref{fig:A-schematic-diagram}).
In sec. \ref{sub:The-response-function} we will show how the optical
response $\Delta R/R$ can be related to the order parameter in superconductors
and CDW systems. 
\begin{figure}
\includegraphics[width=1\columnwidth]{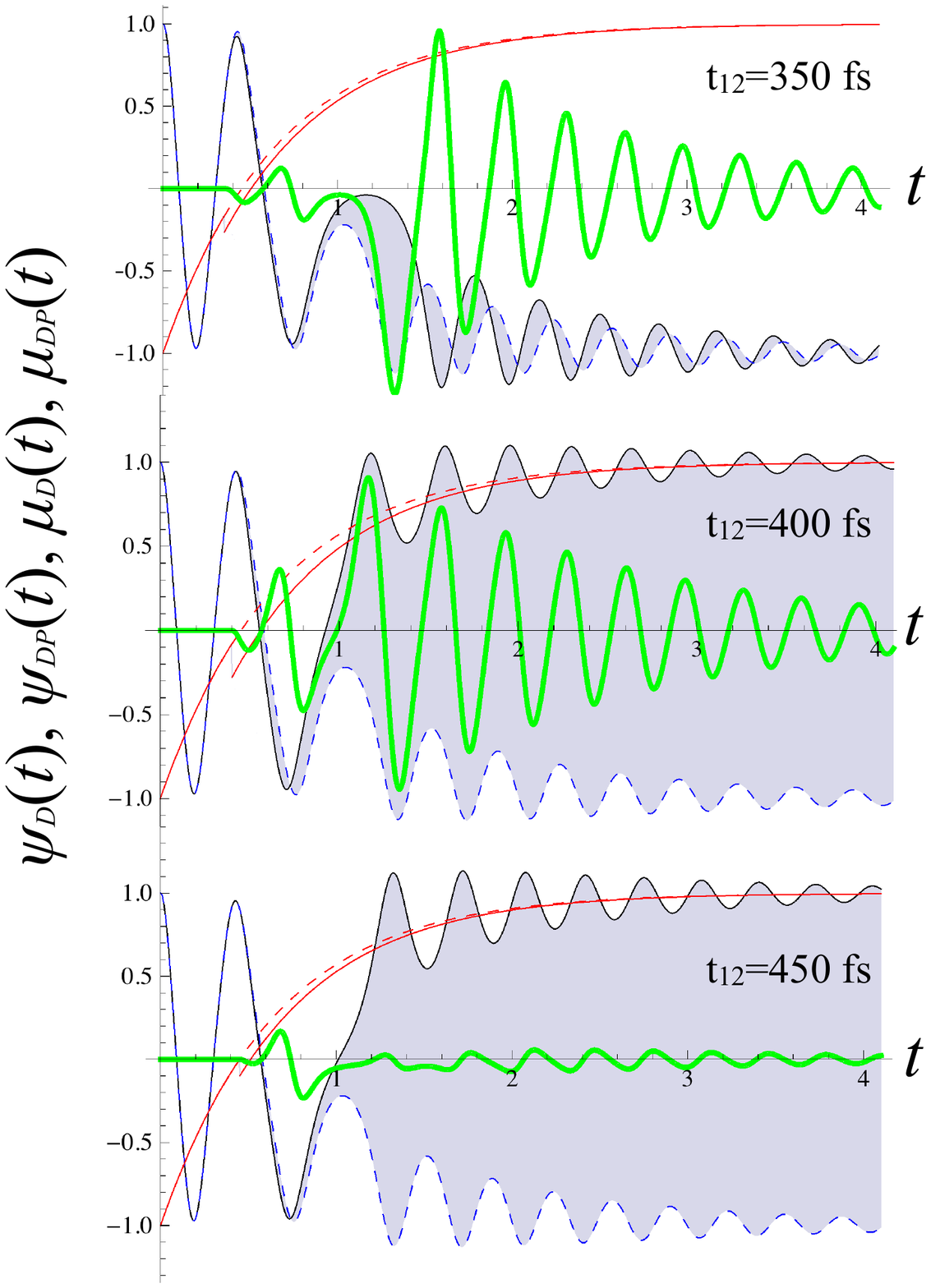}

\caption{The calculated evolution of the order parameter $\psi_{D}(t)$$ $
through the transition after an intense destruction \emph{D} pulse
at $t=0$ (dashed blue line). When an additional \emph{P} pulse is
present, the trajectory $\psi_{DP}(t)$ (solid blue line) can be drastically
modified, depending critically on $\Delta t_{1-2}$. The difference
between the two is highlighted by the shaded areas. The evolution
is calculated for three different time delays $\Delta t_{1-2}=350,400$
and 450 fs. The intensity of the \emph{P} pulse is 10\% of the intensity
of the \emph{D} pulse. The calculation in this figure is done with
an exponential perturbation function $\mu(t)$ for the $D$ pulse
(shown by the dashed red line) and $\mu_{DP}(t)$ with both pulses
(solid red line). The green line is the response function in reflectivity
predicted according to Eq. \ref{eq:responseCDW}.\label{fig:The-calculated-trajectory}}
\end{figure}

On the level of field theory, the trajectory of the system through
the SBT can be modeled within time-dependent Ginzburg Landau theory
with a free energy functional corresponding to the symmetry of the
problem in hand. For a charge density wave system, to capture the
salient physics, we can assume an order parameter $\Psi=A(t,\mathbf{r})e^{i\phi}$
and make the assumption that the relaxation on phase $\phi$ is slow
compared to $\tau_{GL}$, which means that the trajectory is dominated
by the dynamics of \emph{A(t,r)} \cite{Yusupov,Yusupov2}:

\begin{equation}
\frac{\partial^{2}A}{\partial t^{2}}+\Gamma\omega_{AM}\frac{\partial A}{\partial t}-\alpha(t,\mbox{r})\omega_{AM}^{2}A+\omega_{AM}^{2}A^{3}-\xi^{2}\omega_{AM}^{2}\frac{\partial^{2}A}{\partial z^{2}}=0,\label{eq:Equation of motion for CDW}
\end{equation}

where $\omega_{AM}$ is the angular frequency of the collective amplitude
mode, $\xi$ is the coherence length, $\Gamma=\Delta\omega_{AM}/\omega_{AM}$
is the damping of the AM, and $\alpha$$(t,\mathbf{r})$ is temporally
and spatially dependent and is derived from the usual GL temperature
$\alpha(t,\mathbf{r})=1-\mu(t,\mathbf{r})$ describes the control
parameter. $\mu(t,\mathbf{r})$ may be $T(t,\mathbf{r})/T_{c}$, as
given by original GL theory, or may be approximated by an exponential
function which signifies the cooling of the system. The spatial variation
of the light intensity is accounted for by an excitation function
$\mu(t,z)=\frac{T_{e}(t,z)}{T_{c}}\exp(-z/\lambda)$ where $\lambda$
is the optical penetration depth. Using experimental values for $\nu_{AM}=\omega_{AM}/2\pi=2.4$
THz, the line-width $\Delta\nu_{AM}=0.05$ THz and coherence length
$\xi=1$ nm \cite{Yusupov} we can compute $A(t,z)$.

In 3-pulse experiments, the P pulse presents an additional perturbation
which can modify the trajectory, especially if it occurs close to
the critical time $t_{c}$. In Fig. \ref{fig:The-calculated-trajectory}
we show calculated trajectories of $\psi$ for three different delay
times $\Delta t_{1-2}$ of the P pulse with respect to the D pulse.
The ``butterfly effect'' is very evident: $\psi$ may end up as
either +1 or -1, critically depending on $\Delta t_{1-2}$.

The equation of motion for a superconductor within the time-dependent
Ginzburg-Landau (TDGL) model is more complicated because the excitations
are charged and $\psi$ is coupled to the electro-magnetic field,
and are usually considered to be overdamped, so they do not include
any second-order time-derivative: 
\begin{equation}
u\left(\frac{\partial\psi}{\partial t}+i\Phi\psi\right)=-\alpha_{r}(t,\mathbf{r})\psi-\psi|\psi|^{2}-(i\nabla+\mathbf{a})^{2}\psi+\eta\label{eq:tdgl1}
\end{equation}
 
\begin{equation}
\nabla^{2}\Phi=-\nabla\Bigl[\frac{{i}}{{2}}(\psi^{*}\nabla\psi-\psi\nabla\psi^{*})+\mathbf{a}|\psi|^{2}\Bigr].\label{tdgl2-1}
\end{equation}
These equations have been used for analytic and numeric analyses of
dynamics of topological defects (vortices and phase slips) in superconducting
wires, as reviewed in \cite{IK,ludac}. The vector potential $\mathbf{a}$
is written in units of $\frac{\phi_{0}}{2\pi\xi}$ ($\phi_{0}$ is
the flux quantum, $\xi$ is the coherence length) and the electrostatic
potential $\Phi$ in units of $\frac{\hbar}{2e\tau_{\theta}}$, with
$e$ being the elementary charge, $\tau_{\theta}$ is the transverse
relaxation time and $\hbar$ the reduced Planck constant. The only
dimensionless parameter left in the equation is the ratio $u=\frac{\tau_{\rho}}{\tau_{\theta}}$
between the longitudinal relaxation time and the transverse relaxation
time. Typically, for high-temperature superconductors $\tau_{\theta}\approx1\mbox{ ps}$
and $u\approx5$ \cite{ludac}. The Langevin noise term $\eta$ introduces
microscopic fluctuations into the dynamics \cite{ludac}.

\section{Control of the quench rate and optical response function\label{sec:Control-of-the}}

In the simplest approximation, suggested by the usual $\alpha$ parameter
in the GL equations, the excitation function $\mu$ can be the system
temperature. In the case where the electronic system is being discussed,
such as superconductors and CDWs, it is the electronic temperature
$T_{e}(t)$ which is the relevant control parameter. In recent years,
significant progress has been made in understanding the time evolution
of electron and lattice temperatures in superconductors and CDWs following
excitation by a laser pulse. The quench rate is determined by the
energy relaxation rate of electrons at early times up to $\sim$1
ps, and the lattice $T_{L}(t)$ on longer timescales (tens of picoseconds
and beyond). The time-evolution of $T_{e}$$(t)$ and $T_{\mathrm{L}}(t)$
respectively, which govern the quench process, may be estimated using
a two-temperature model \cite{Allen}: $\gamma_{\mathrm{e}}T_{\mathrm{e}}\frac{dT_{\mathrm{e}}}{dt}=-\gamma_{\mathrm{L}}(T_{\mathrm{e}}-T_{\mathrm{L}})+E(t)$
and $C_{\mathrm{L}}(T)\frac{dT_{\mathrm{L}}}{dt}=\gamma_{\mathrm{L}}(T_{\mathrm{e}}-T_{\mathrm{L}}),$
where $E(t)$ is a function (Gaussian) describing the energy density
per unit time supplied by the laser pulse. The pulse energy $E_{D}=E_{p}/d^{2}$
where $E_{p}$ is the pulse energy. We used the temperature-dependent
thermal constants $\gamma_{\mathrm{e}}(T)$ and $C_{\mathrm{L}}(T)$
from experimental data\cite{C} and the measured energy loss rate
$\gamma_{\mathrm{L}}\simeq340$ K/ps for La$_{1.9}$Sr$_{0.1}$CuO$_{4}$
\cite{Gadermaier}. The dependence of $T_{\mathrm{e}}(t)$ and $T_{\mathrm{L}}(t)$
on the energy density per pulse $U=\int E(t)dt$ is shown in Fig.
\ref{fig:The-calculated-evolution}a) for the first few picoseconds.
The red contour lines show when either $T_{\mathrm{e}}$ or $T_{\mathrm{L}}$
crosses $T_{\mathrm{c}}$. Initially, the rate of cooling $\gamma_{e}=\frac{dT_{e}}{dt}$
is very rapid until $T_{e}$ reaches $T_{L}$. Thereafter, $T_{e}$
follows $T_{L}$, and is dominated by phonon anharmonic decay or phonon
escape and thermal diffusion processes, all of which are much slower
than electronic energy loss. Thus, to change the quench rate we can
adjust the energy density $U_{D}$ of the \emph{D} pulses, so that
the system either cools rapidly through electron thermalization, or
more slowly by phonon decay and diffusion processes. As shown in Fig.
\ref{fig:The-calculated-evolution}a), for low \emph{$U_{D}$}, only
$T_{e}$ exceeds $T_{c}$, so the quench back through $T_{c}$ will
be fast and dominated by electronic cooling. For larger \emph{$U_{D}$,}
$T_{L}$ exceeds $T_{c}$, so on cooling the quench rate though $T_{c}$
is significantly slower and dominated by lattice cooling.

\subsection{The response function\label{sub:The-response-function}}

The key issue is detection of the order parameter through the SBT.
Here we show how for superconductors and CDW systems, the response
measured in 3-pulse experiments is related to the order parameter.
While a full kinetic theory which would take into account single particle
and phonon relaxation processes through the transition is beyond reach
at present, we can obtain good approximations for the quasiparticle
response functions in superconductors and collective mode in CDW systems
respectively based on well-accepted phenomenological theory. 

The temperature dependence of the difference of the optical reflectivity
between the superconducting state $R_{s}$ and the normal state $R_{n}$
in a superconductor has been found to be described very well by an
expression derived from the Mattis-Bardeen formula \cite{Kusar}: 

\begin{equation}
R_{s}(T)-R_{n}=A\frac{\Delta\left(T\right){}^{2}}{\omega^{2}}ln\left[\frac{1.47\omega}{\Delta\left(T\right)}\right]
\end{equation}

where $\Delta(T)=1-\left(T/T_{c}\right)^{2}$ is usually used to describe
the temperature dependence of the gap, $\omega$ is the frequency
of light in units of $\hbar$, $R_{n}$ is the reflectivity in the
normal state. We need to calculate the transient change of reflectivity
$\delta R$, as the system is evolving through the transition in time,
so we explicitly replace $\Delta(T)$ with $\Delta(t)$. Using the
2-fluid model, we substitute $\Delta^{2}=\Delta_{0}^{2}n_{s}=\Delta_{0}^{2}(1-n_{q})$,
where $n_{s}$ is the superfluid density and $n_{q}$ is the quasiparticle
density, and $\Delta_{0}$ is the gap at $T=0$. We can relate $\delta R$
to the photo-excited carrier density $n_{p}$ using the fact that
$\delta n_{q}=n_{p}$:

\begin{equation}
\delta R=\frac{\partial R}{\partial\Delta}\frac{\partial\Delta}{\partial n_{q}}\delta n_{q}=\frac{A\Delta_{0}^{2}}{2\omega^{2}}\left(1-2ln\left[\frac{1.47\omega}{\Delta\left(t\right)}\right]\right)n_{p}\label{eq:transient reflectivity response}
\end{equation}

If we ignore the derivative of the logarithmic correction with respect
to $\Delta$, substituting $\Delta_{0}$ for $\Delta(t$) in the logarithm,
we obtain $R_{s}(t)-R_{n}\simeq A\frac{\Delta\left(t\right){}^{2}}{\omega^{2}}ln\left[\frac{1.47\omega}{\Delta_{0}}\right]$.
$\delta R$ then simplifies to: 
\begin{equation}
\delta R=\frac{\partial R}{\partial\Delta}\frac{\partial\Delta}{\partial n_{q}}\delta n_{q}=\frac{2A\text{\ensuremath{\Delta(t)}}}{\omega^{2}}\delta\Delta\propto-\frac{A\text{\ensuremath{\Delta_{0}^{2}}}}{\omega^{2}}n_{p}.\label{eq:simple response}
\end{equation}

Under near-bottleneck conditions, when the QPs and the phonons are
in near-equilibrium, $n_{p}$ is given by \cite{Kabanov}:

\begin{equation}
n_{p}=\frac{1/(\Delta(T(t)))+T(t)/2)}{1+B\sqrt{2T(t)/(\Delta(T(t))}\text{Exp}[-\Delta(T(t))/T(t)]}.\label{eq:epsilon in terms of psi}
\end{equation}

where $\Delta$ is the superconducting gap, $B=\nu/N(0)\hbar\Omega_{c}$,
where $\nu$ is the effective number of phonon modes of frequency
$\hbar\Omega_{c}$ per unit cell participating in the relaxation process.
Here we have explicitly written the temperature $T$ to be time dependent.
$B$ can be determined by fitting the temperature dependence of $\Delta R/R$
and $N(0)$ is the density of electronic states at the Fermi energy.
Substituting $\Delta_{0}\psi(t)$ for $\Delta(t)$ from the solution
of Eq. \ref{tdgl2-1}, we obtain:

\begin{equation}
\delta R/R\propto\frac{A\Delta_{0}^{2}}{2\omega^{2}}\left(1-2ln\left[\frac{1.47\omega}{\Delta\left(t\right)}\right]\right)\frac{1/(\Delta_{0}|\psi(t)|+T(t)/2)}{1+B\sqrt{2T(t)/\Delta_{0}|\psi(t)|}\text{Exp}[-\Delta_{0}|\psi(t)|/T(t)]}\label{eq:response}
\end{equation}

\begin{figure}
\includegraphics[width=12cm]{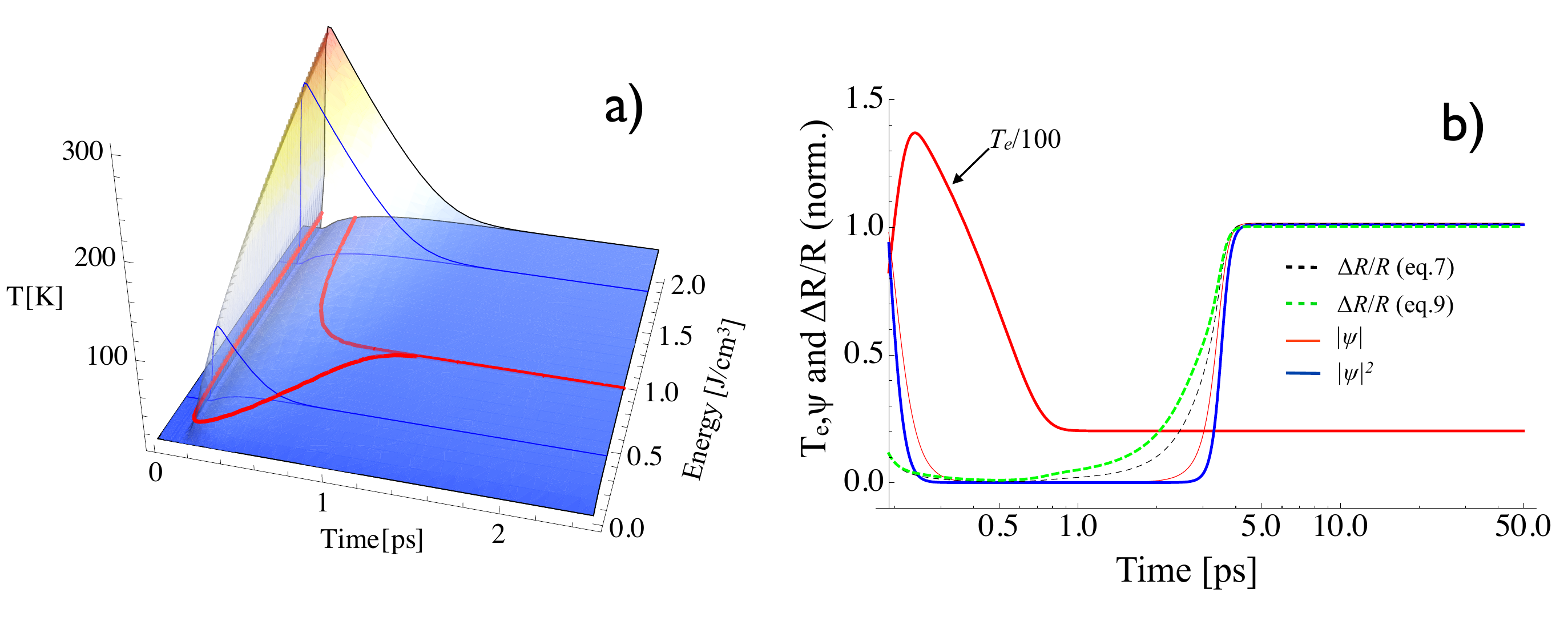}

\caption{a) The calculated evolution of $T_{e}$, $T_{L}$ as a function of
$D$ pulse energy. b) The order parameter $|\psi|$ and reflectivity
response $\Delta R/R$ as a function of time through the quench. The
calculation of $T_{e}$ and $T_{L}$ is performed using the two-temperature
model \cite{KA,Allen} using the appropriate specific heat and thermal
conductivity functions for La$_{1.9}$Sr$_{0.1}$CuO$_{4}$ from experimental
data \cite{Kusar}. The response functions compare the response calculated
with Eq. \ref{eq:response} and \ref{eq:simple response} with $|\psi|$
and $|\psi|^{2}$. \label{fig:The-calculated-evolution}}
\end{figure}

The response functions using the appropriate constants for La$_{1.9}$Sr$_{0.1}$CuO$_{4}$
are plotted in Fig. \ref{fig:The-calculated-evolution}, together
with the time evolution of $T_{e}(t)$ for $E=1.6$ J/cm$^{3}$, and
the order parameter $|\psi(t)|$ and $|\psi(t)|^{2}$ for the case
of a homogeneous superconductor calculated using Eq. \ref{eq:tdgl1}
in the absence of fieled $\mathbf{a=0}$, and $\Phi=0$. The experimental
reflectivity response $\delta R/R$ is close to the square of the
order parameter $|\psi|^{2}$ over a large range of $t$, particularly
for large $\psi$. The calculated case for an inhomogeneous superconductor
which takes into account the depth profile of the laser beam is discussed
in the next section where it is compared with experimental data.

In CDWs, when discussing the transient response of collective amplitude
mode, the response is related to the displacement $\Delta R$ of the
atoms from equilibrium position $R_{0}$, where $\Delta R\propto\psi-\psi_{0}$,
where $\psi_{0}$ is the equilibrium value of the order parameter.
To obtain the optical response, we can expand the dielectric constant
near CDW phase transition in powers of the order parameter \cite{Ginzburg}:
\begin{equation}
\epsilon=\epsilon_{0}+c_{2}|\psi|{}^{2}.\label{eq:2}
\end{equation}
 Here $\epsilon_{0}$ is the dielectric constant of the high temperature
symmetric phase, $c_{2}$ is a real constant.

The relevant response in \emph{D-P-p} experiments can be expressed
in terms of the difference between the response \emph{with} and \emph{without}
the $P$ pulse, proportional to $\Delta|\psi|^{2}$, so to first order
\cite{Yusupov}:

\begin{equation}
\frac{\Delta R}{R}\simeq\frac{\Delta\epsilon}{\epsilon}\propto\psi_{DP}^{2}(t,\mathbf{r},\Delta t_{12})-\psi_{D}^{2}(t,\mathbf{r}).\label{eq:responseCDW}
\end{equation}

This response has been tested for the case of the CDW transition in
TbTe$_{3}$ \cite{Yusupov}. We see that by careful design of the
experimental probe and the use of 3 pulse techniques, it is possible
to identify and even isolate the dominant contribution to the optical
response which is related to the order parameter, thus providing valuable
information of the temporal evolution of $\psi$ by measurement of
either single particle and collective excitations through the SBT. 

\section{Superconductors: probing vortex dynamics on the picosecond timescale.\label{sec:Superconductors:-probing-vortex}}

\begin{figure}
\includegraphics[width=1\columnwidth]{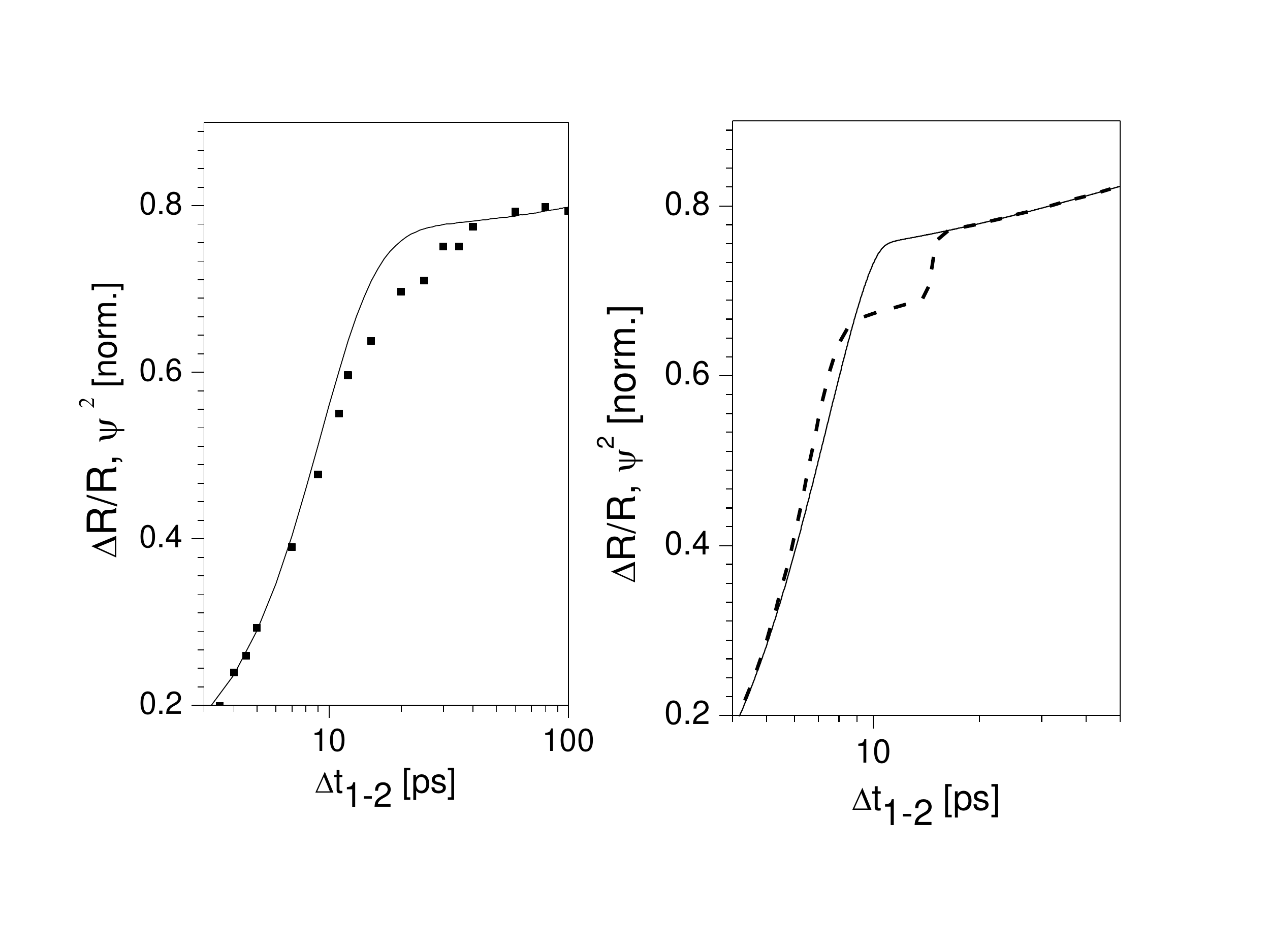}

\caption{The recovery of the order parameter $\psi$ after an ultrafast quench
in La$_{1.9}$Sr$_{0.1}$CuO$_{4}$. The data points (left panel)
represent the dependence of the maximum $\Delta R/R$ response as
a function of $\Delta t_{1-2}$, while the solid curve is the trajectory
of $|\psi(t)|^{2}$ calculated from Eq. \ref{tdgl2-1}, normalized
to the value at long times after the \emph{D} pulse. The right panel
shows the calculated trajectory when fluctuations are included in
Eq. \ref{tdgl2-1} via the $\eta$ term (dashed line). The trajectory
without noise $\eta$ is included for reference (black line) for the
same parameters. The difference between the two is attributed to the
effect of vortices spontaneously created via the KZ mechanism, which
annihilate on a timescale of 20 ps according to theory predictions.
A comparison with the data suggests the formation of vortices in La$_{1.9}$Sr$_{0.1}$CuO$_{4}$
created in the quench on a timescale between 10 and 30 ps.\label{fig:SC recovery}
Note that we have limited ourselves to the time interval where $\psi\rightarrow\psi_{equilibrium}$,
so $\Delta R/R\propto|\psi|$ is a valid approximation (see Fig. \ref{fig:The-calculated-evolution}b)). }
\end{figure}

In Figure \ref{fig:SC recovery}a) we show the time-evolution of $\delta R/R$
in La$_{1.9}$Sr$_{0.1}$CuO$_{4}$ measured with a $P-p$ sequence
at a sample base temperature of 4 K after a 50 fs laser pulse. All
laser pulses have a wavelength of 800 nm. The \emph{D} pulse energy
$E_{\mathrm{D}}=0.8$ J/cm$^{3}$ is adjusted to heat the sample above
$T_{c}$. (The threshold values for the destruction of the SC state
were previously reported in \cite{Kusar}). 

According to Eq. \ref{eq:epsilon in terms of psi}, the QP amplitude
in \emph{P-p} experiments are related to $|\psi|$, which allows us
to compare the data with the calculated trajectory based on a solution
of Eqs. \ref{eq:tdgl1} and \ref{tdgl2-1}. The numerical solution
calculated without the fluctuation term $\eta$ is shown by the solid
line in Fig. \ref{fig:SC recovery}a). Comparing with the measured
data, we see that the prediction is remarkably good overall, except
for a depression of the order parameter of approximately 10-20 \%
between 10 and 30 ps. This discrepancy may be attributed to vortices
which form as a result of the fast quench through $T_{c}$. $\psi\rightarrow0$
in the vortex cores, so the spatially averaged value of $\psi$ will
be depressed if they are present. To try and model the effect of vortices
we compare calculations with and without the fluctuation term $\eta$,
while keeping all other parameters unchanged. The introduction of
fluctuations gives rise to a depression of $|\psi|$ in the region
around 10 ps, shown by the dashed line in Fig. \ref{fig:SC recovery}b).
Qualitatively, we see that the presence of vortices may explain the
depression of the order parameter around 10 ps in La$_{1.9}$Sr$_{0.1}$CuO$_{4}$.
Considering the simplicity of both the TDGL model and the crudeness
of the approximations for the response function (Eq. \ref{eq:epsilon in terms of psi}),
the agreement between data and theory is quite reasonable.

\section{Topological defects in CDWs: coherent dynamics.\label{sec:Topological-defects-in} }

\begin{figure}
\includegraphics[width=1\columnwidth]{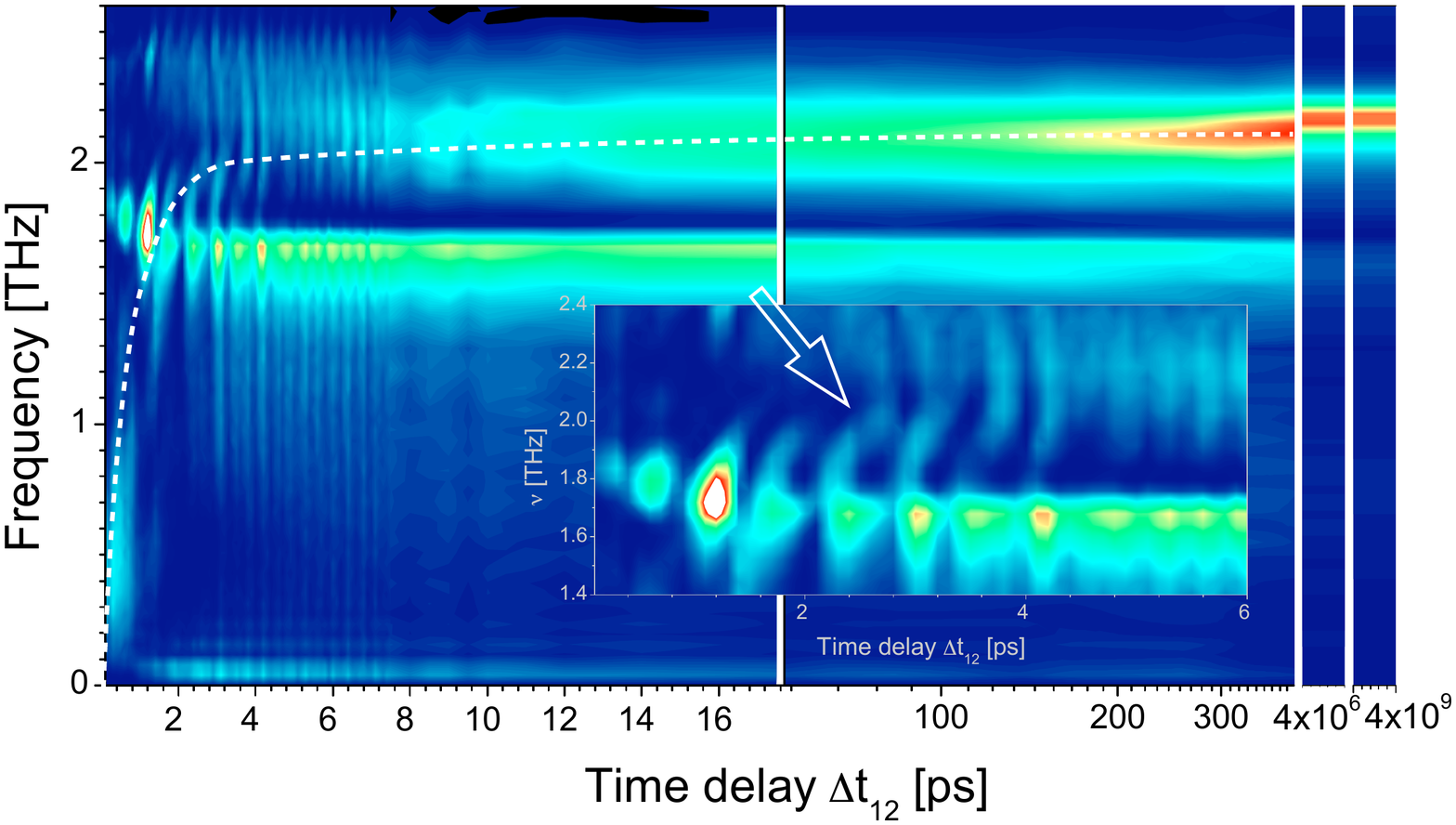}

\caption{The temporal evolution of the collective mode spectrum in DyTe$_{3}$
through the quench into the ordered phase. The plot shows the amplitude
of the spectrum obtained from the transient reflectivity $\Delta R/R$
as a function of $\Delta t_{2-3}$ for different $\Delta t_{1-2}$.
The AM shows some intensity around 0.5 THz at short times, then crosses
a phonon at 1.68 THz at around 2 ps and eventually settles down to
2.2 THz at long times. The dotted line shows the recovery of the electronic
single particle gap determined from the recovery of the SP peak scaled
onto the $\omega-\Delta t_{1-2}$ plot. At early times, coherent oscillations
of the order parameter cause fluctuations of the intensity of the
AM and the 1.68 THz phonon. Interference between the phonon and the
AM is clearly observed, including a renormalization of the phonon
frequency from $\sim$ 1.85 THz at early times (in the high-symmetry
state) to 1.68 THz in the ordered state. The insert shows the dynamics
at short times. Distortions of the line-shapes in time are attributed
to the presence of topological defects created in the quench to the
CDW state (see arrow) on the basis of comparisons between theory and
experiment (see Fig. \ref{fig:Calculated OP and response}).\label{fig:CDW}
The sample base temperature is 4 K throughout the experiment. The
color scale is normalized, so that the maximum amplitude is 1 (red).
(In the insert, the scale is expanded to show the additional spectral
features shown by the arrow.)}
\end{figure}

Transition metal tri-tellurides such as DyTe$_{3}$ are excellent
systems for femtosecond measurements of system trajectories through
SBTs to a CDW state. Recently, the evolution of single particle and
collective excitations though the CDW transition have been reported
in TbTe$_{3}$ \cite{Yusupov}. Here we show the temporal evolution
of the order parameter in a related material, DyTe$_{3}$, which undergoes
a transition to a CDW state at 305K. The temporal evolution of the
collective mode spectrum after the \emph{D} pulse is shown in Fig.
\ref{fig:CDW}. The spectrum was obtained from the Fourier transform
of the transient reflectivity $\Delta$R/R recorded at each $\Delta t_{1-2}$
as a function of $\Delta t_{2-3}.$ We identify two modes in the spectra
at $\sim$1.68 THz and $\sim$ 2.2 THz at long times $\Delta t_{1-2}>10$
ps. The latter is assigned to the amplitude mode (AM) of the CDW based
on its distinct temperature dependence \cite{Yusupov2-1}, while the
former is a phonon mode (PM). At $t\simeq1.3$ ps, the AM exhibits
mixing with the PM. Oscillations of the order parameter are clearly
visible at short times and appear as modulations of the intensity
of both modes occurring up to $\sim$ 8 ps. At longer times, the intensity
oscillations and the frequency recovers to 2.2 THz after $4\times10^{6}$
ps. No further relaxation takes place beyond this timescale. 

\begin{figure}
\includegraphics[height=6cm]{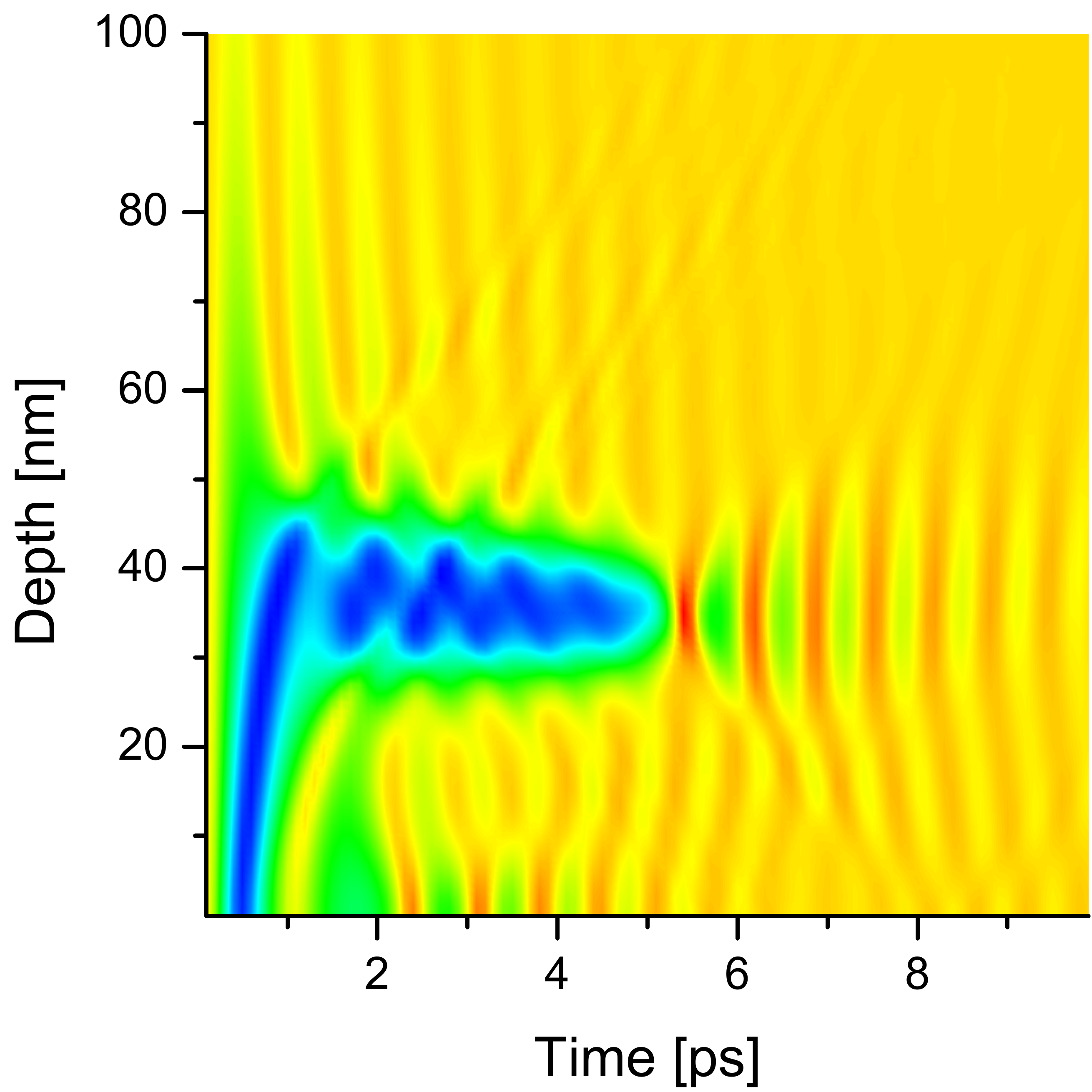}\includegraphics[height=6cm]{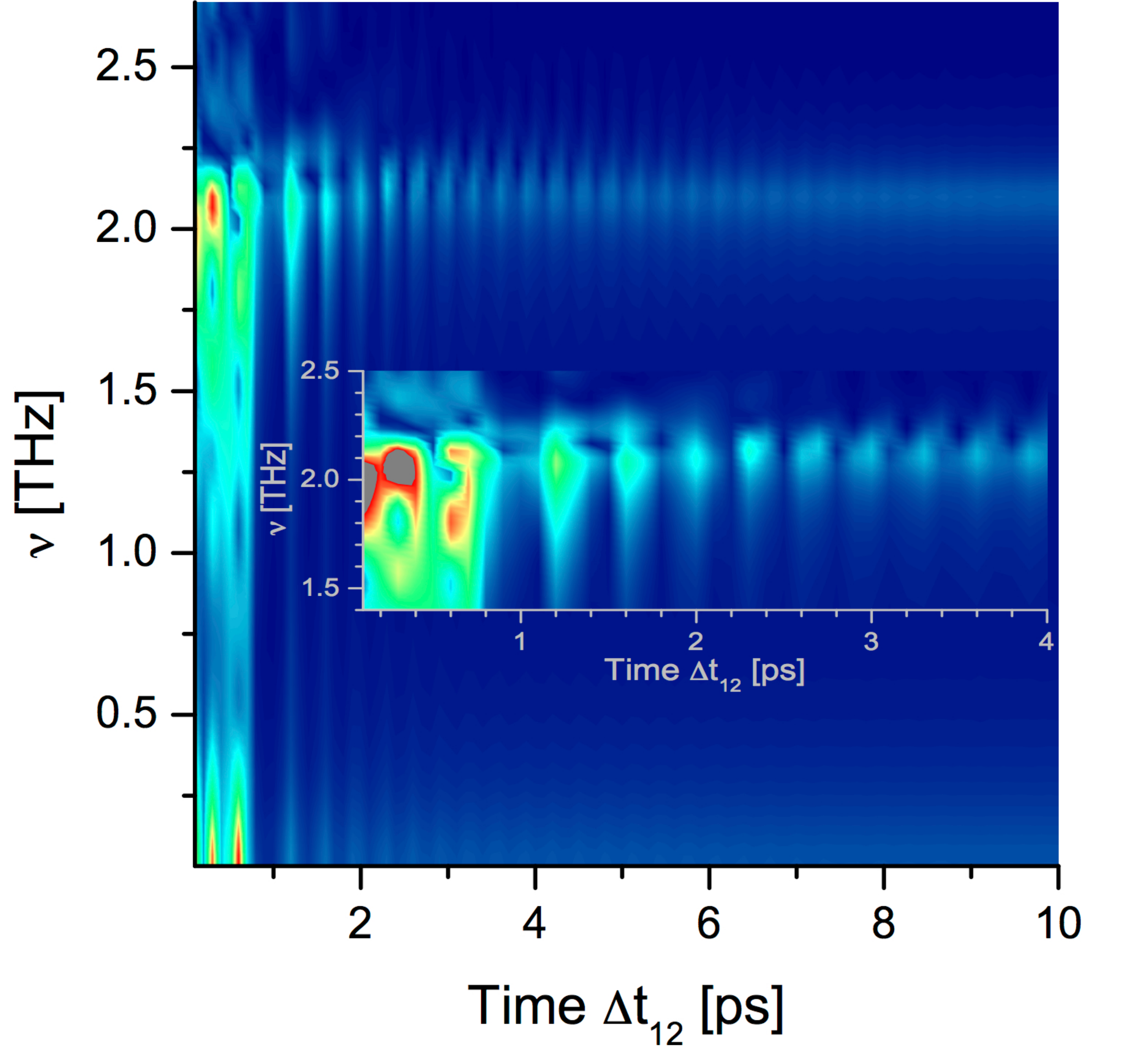}

\caption{a) The calculated order parameter (\foreignlanguage{british}{colour}
scale) as a function of time and depth into the sample after a laser
pulse as a solution to Eq. \ref{eq:Equation of motion for CDW} and
parameters for DyTe$_{3}$. b) The intensity oscillations are calculated
from Eq. \ref{eq:responseCDW}, taking into account the appropriate
boundary conditions and inhomogeneity of the laser field in Eq. \ref{eq:Equation of motion for CDW}.
Distortions to the line-shapes are attributed to the annihilation
of topological defects, similarly as observed in TbTe$_{3}$ \cite{Yusupov}.
The color scale is normalized, so that the maximum amplitude is 1
(red) and minimum is -1 (blue). (In the insert, the color scale is
expanded to show the additional spectral features) \label{fig:Calculated OP and response}}
\end{figure}

In Fig. \ref{fig:Calculated OP and response} a) we show the calculated
$\psi$$(z,t)$ from Eq. \ref{eq:Equation of motion for CDW} using
parameters for DyTe$_{3}$: $\omega_{AM}=2.2$ THz, the AM linewidth
$\Gamma=0.3$ THz, and coherence length $\xi=1.3$ nm from \cite{xi}.
For simplicity, the driving term $\alpha(t,z)$ in Eq. \ref{eq:Equation of motion for CDW}
is assumed to be exponential in time and in space: $\alpha=exp(-t/\tau-z/\lambda)$,
where the penetration depth is given by the experimental value $\lambda=20$
nm \cite{lambda}. The calculation predicts the formation of three
domains created within the first picosecond, parallel to the surface,
with domain walls at $\sim$ 30 nm and $\sim$ 45 nm. After 5 ps,
the domains walls are annihilated, leaving behind a single domain.
The event is accompanied by the emission of a finite frequency dispersive
amplitude waves traveling towards the surface and into the sample.
The velocity of the amplitude wave is seen from Fig. \ref{fig:Calculated OP and response}
to be approximately $v_{A}\simeq10$ nm/ps. The wave reaches the surface
with 6-8 ps causing a disturbance of the AM which is visible as a
temporal distortion of the AM spectral line-shape. (The insert to
Fig. \ref{fig:CDW} highlights the distortions of $\omega-\Delta t_{12}$
spectra). The calculated spectrum (using Eq. \ref{eq:responseCDW})
is shown in Fig. b). The predicted distortions have a great deal of
similarity with the experimentally observed evolution of the spectrum
shown in Fig. \ref{fig:CDW}. Unfortunately the presence of the phonon
interference in this material complicates the detailed comparison
between theory and experiment. Similar, although more pronounced distortions
were observed in TbTe$_{3}$, where the interference from other phonons
is less problematic \cite{Yusupov}.

In \cite{Mertelj} it is further shown that incoherent intrinsic topological
defect dynamics in the related system TbTe$_{3}$ occurs on a timescale
of $\sim$ 30 ps, similar to the timescale of vortex annihilation
dynamics inferred from the experiments on La$_{1.9}$Sr$_{0.1}$CuO$_{4}$.
The experiments show that pinned-defect-related annihilation appears
to be present on much longer timescales of $10^{-10}-10^{-6}$ s.

\section{Discussion}

We have demonstrated that femtosecond optical experiments with multiple
pulse techniques open the way to detailed studies of the temporal
dynamics of systems undergoing SBTs and can be interpreted in the
context of a phenomenological field theory without resort to microscopic
theory. The basic analogy with cosmology and elementary particle collisions
comes from the fact that the SBT take place in temporally evolving
systems with high temperature intial conditions. Single particle fermionic
and collective bosonic excitations can be unambiguously identified
and related to the temporal evolution of the order parameter through
the transition and topological defects (domain walls and vortices)
are shown to lead to experimentally observable phenomena. It appears
that CDWs are eminently more suitable for investigation of topological
defects than superconductors because the collective mode conveys significantly
more information of the system trajectory than single-particle excitations,
particularly relating to coherent defect dynamics. We have shown that
by adjusting the laser excitation energy density, the quench rate
can be varied and controlled, leading to the possibility of investigating
domain wall recombination. Further studies as a function of quench
rate and temperature may be expected to reveal systematic \foreignlanguage{british}{behaviour},
which can be related to the predictions of the KZ mechanism for the
generation of topological defects. In the context of the discussion
regarding the possible cosmological importance of creation and annihilation
of cosmic domains and strings \cite{Kibble}, the present paper demonstrates
that \emph{coherent} creation and annihilation of\emph{ }topological
defects leads to observable consequences in condensed matter systems.
Contrary to previous experiments, the distinction between coherent
and incoherent recombination dynamics, as well as intrinsic and extrinsic
defect-related annihilation is quite clear, because the rates of decay
are vastly different: coherent domain recombination occurs on a $\tau_{coh}\simeq$
3 ps timescale, while intrinsic incoherent domain wall recombination
timescale is $\tau_{i}\simeq$ 30 ps. Finally, extrinsic defect-related
dynamics occurs on timescales of $\tau_{ext}\sim$ 100 ps or longer
\cite{Mertelj}. The experiments thus establish a hierarchy of timescales
associated with the annihilation of topological defects.

The evolution of the single particle response as a function of time
after destruction by a laser pulse in the La$_{1.9}$Sr$_{0.1}$CuO$_{4}$
superconductor suggests that a depression of the order parameter compared
with the theory prediction can be understood if vortices are included
in the within an inhomogeneous model, suggesting intrinsic vortex
creation and annihilation in La$_{1.9}$Sr$_{0.1}$CuO$_{4}$ takes
place on a timescale of 10-20 ps. Although one might argue that the
discrepancy between the trajectory of $|\psi|$ and the measured response
may be attributed to breakdown of the approximations made in deriving
the response function Eq. \ref{eq:epsilon in terms of psi}, these
problems are not expected to be important in the region around 10-20
ps, but rather at shorter times, closer to $t_{c}$: when $\psi\rightarrow1$,
the gap is fully developed, and the QP response is expected to be
described well by the \foreignlanguage{british}{Rothwarf-Taylor} model
in the bottleneck regime\cite{Rothwarf}, where the gap $\Delta(t)\rightarrow1$
and is almost constant, so Eq. \ref{eq:epsilon in terms of psi} is
reasonably valid, as shown in Fig. \ref{fig:The-calculated-evolution}. 

We point out that in the context of more general field theories of
related time-evolving systems, apart from the AM, two other bosons
observed in CDW systems are of interest. The first is created as a
result of the annihilation of domain walls shown in Fig. 6. Upon annihilation,
two finite-frequency (massive) $\psi-$waves are emitted which propagate
perpendicular to the surface into the sample and towards the surface.
Calculations of the ``wave effect'' on the temporal evolution of
the AM spectral shape appear to be confirmed by experiments in DyTe$_{3}$
(discussed here) and in TbTe$_{3}$ \cite{Yusupov}.

The second bosonic excitation of interest is the phonon mode (PM)
with an equilibrium frequency of $\omega_{PM}=$1.68 THz. The time-evolution
of this mode is very different from the evolution of the AM: it does
not show critical behaviour near $t_{c}$. Instead its frequency is
just slightly renormalized in the low symmetry state. Before $t_{c}$,
the PM frequency is $\omega_{PM}=$$1.85\pm0.02$ THz. Because its
frequency $\omega_{PM}$ in the ordered state is lower than $\omega_{AM}$
($=2.2$ THz), it crosses the AM as the latter hardens after $t_{c}$
as indicated in Fig. 5. At $\Delta t_{12}\simeq1.3$ ps the AM interferes
with the PM, and the phonon frequency is renormalized for $\Delta t_{12}>2$
ps, to $\omega_{AM}=$ $1.68\pm0.02$ THz. 

In experiments where the control parameter is temperature $T$ rather
than time, this behaviour is well understood \cite{Yusupov2-1}: below
the critical temperature $T_{c}$, as the AM hardens with decreasing
$T$, it crosses the PM at some intermediate temperature, displaying
mode mixing. The coupling of the AM and the PM in compounds such as
DyTe$_{3}$$ $ is relatively weak, the off-diagonal matrix element
being $\delta\simeq0.1\pm0.01$ THz \cite{Yusupov2-1}. The PMs form
a large reservoir of excitations which are weakly coupled to the order
parameter. The symmetry of the PM plays an important role in the temporal
behaviour. By definition, the order parameter is totally symmetric,
(\emph{A} representation) so even-symmetry modes are expected to couple
strongly to it. When the system has inversion symmetry, such as here,
(DyTe$_{3}$ has $D_{2h}$ point group symmetry) the odd-symmetry
modes do not couple to the $\psi$ to first order. Odd symmetry modes
thus display some characteristics of weakly interacting massive particles
in cosmology. The observed PM has a large mass (frequency) and is
weakly coupled to the order parameter, either because they have inappropriate
symmetry (in which case the weak coupling comes from higher order
interactions), or the matrix element coupling to A symmetry excitations
is small. In contrast to the AM and the fermionic excitations which
emerge after the SBT takes place, the PM excitations exist \emph{before}
the SBT (i.e. before and after the Big Bang within the cosmological
paradigm). By analogy with the PM, dark matter excitations may be
thought of as a remnant excitations from before $t_{c}$. As such
they are external to the GL (or Standard) model. 

We conclude that building on the analogy between a laser-induced \emph{e-h}
plasma in condensed matter systems and the plasma created in the collision
of elementary particles, or with primordial conditions in the Universe,
gives us a practical laboratory-scale playground for the exploration
of temporally evolving systems though symmetry breaking transitions.
Remarkably, the CDW model system give rise to two kinds of cosmological
analogies, namely of the AM as the analogue to the Higgs boson and
odd-symmetry phonons as dark matter excitations. The CDW systems reveal
some hitherto unexplored bosonic excitations created under nonequilibrium
conditions which may be expected to have observable counterparts in
other temporally evolving systems. The reflectivity response function
derived for a superconductor allows the investigation of the trajectory
of the order parameter through the transition, opening the way to
studies of intrinsic vortex dynamics.

\section{Acknowledgments}

We wish to acknowledge discussions with Prof. T. Kibble on the importance
of varying the quench rate in these experiments. We also acknowledge
the use of DyTe$_{3}$ and LaSrCuO single crystal samples kindly provided
by I. Fischer and S. Sugai respectively.

\section*{References}{}

\end{document}